\documentclass[aps,prl,nofootinbib,amsmath,amssymb,twocolumn,10pt]{revtex4}

\pdfoutput=1

\usepackage{graphicx}
\usepackage{dcolumn}
\usepackage{bm}
\usepackage{amssymb}
\usepackage{latexsym}
\usepackage{booktabs}
\usepackage{amsmath}
\usepackage{multirow}
\usepackage[colorlinks=true, linkcolor=red, citecolor=blue]{hyperref}

\newcommand{\be}{\begin{equation}}
\newcommand{\ee}{\end{equation}}
\newcommand{\bq}{\begin{eqnarray}}
\newcommand{\eq}{\end{eqnarray}}

\begin{document}

\title{Weighing neutrinos in dynamical dark energy models}

\author{Xin Zhang}
\affiliation{Department of Physics, College of Sciences,
Northeastern University, Shenyang 110004, China}
\affiliation{Center for High Energy Physics, Peking University, Beijing 100080, China}

\begin{abstract}

We briefly review the recent results of constraining neutrino mass in dynamical dark energy models using cosmological observations and summarize the findings. (i) In dynamical dark energy models, compared to $\Lambda$CDM, the upper limit of $\sum m_\nu$ can become larger and can also become smaller. In the cases of phantom and early phantom (i.e., the quintom evolving from $w<-1$ to $w>-1$), the constraint on $\sum m_\nu$ becomes looser; but in the cases of quintessence and early quintessence (i.e., the quintom evolving from $w>-1$ to $w<-1$), the constraint on $\sum m_\nu$ becomes tighter. (ii) In the holographic dark energy (HDE) model, the tightest constraint on $\sum m_\nu$, i.e., $\sum m_\nu<0.105$ eV, is obtained, which is almost equal to the lower limit of $\sum m_\nu$ of IH case. Thus, it is of great interest to find that the future neutrino oscillation experiments would potentially offer a possible falsifying scheme for the HDE model. (iii) The mass splitting of neutrinos can influence the cosmological fits. We find that the NH case fits the current observations slightly better than the IH case, although the difference of $\chi^2$ of the two cases is still not significant enough to definitely distinguish the neutrino mass hierarchy.

\end{abstract}

 \maketitle

The discovery of neutrino oscillation indicates that neutrinos have masses and each flavor state is actually a superposition of three mass states with masses $m_1$, $m_2$, and $m_3$. However, the neutrino oscillation experiments are not able to measure the absolute masses of neutrinos, but can only measure the squared mass differences between the neutrino mass eigenstates---the solar and reactor experiments gave $\Delta m_{21}^2\simeq 7.5\times 10^{-5}$, and the atmospheric and accelerator beam experiments gave $|\Delta m_{31}^2|\simeq 2.5\times 10^{-3}$, which indicates that there are two possible mass orders, i.e., the normal hierarchy (NH) with $m_1<m_2\ll m_3$ and the inverted hierarchy (IH) with $m_3\ll m_1<m_2$. 

To work out the absolute masses, one needs at least an additional relation between the three neutrino mass eigenstates. Since massive neutrinos play an important role in the evolution of the universe, they could leave distinct signatures on the cosmic microwave background (CMB) and large-scale structure (LSS) at different epochs of evolution of the universe. Actually, these signatures can be extracted from the available cosmological observations, to a large extent, from which the total mass of neutrinos can be constrained. In recent years, the CMB temperature and polarization power spectra, in combination with LSS and cosmic distance measurements, have been providing more and more tight limits on the total neutrino mass. 

Recently, it was realized that the properties of dark energy could influence the cosmological weighing of neutrinos largely. Although the cosmological constant $\Lambda$ (the model of $\Lambda$ together with cold dark matter is usually called the $\Lambda$CDM model) can explain the various observations quite well, many other dark energy candidates are not yet excluded by the current observations. In fact, some dynamical dark energy models are still rather competitive in fitting the current observational data. Therefore, the impacts of dark energy on weighing neutrinos should be seriously investigated. 

In a recent work \cite{Zhang:2015uhk}, I carefully investigated how the dark energy properties impact the cosmological limits on the total neutrino mass $\sum m_\nu$. In this study, as typical examples, only two simplest dynamical dark energy models that have only one more parameter compared to $\Lambda$CDM are considered, i.e., the $w$CDM model (in which dark energy has a constant equation-of-state parameter $w$) and the holographic dark energy (HDE) model (that has the only additional parameter $c$ in the definition of its energy density, $\rho_{\rm de}=3c^2 M_{\rm pl}^2 R_{\rm EH}^2$, where $M_{\rm pl}^2$ denotes the reduced Planck mass and $R_{\rm EH}$ the event horizon size of the universe; note that here $c$ is a dimensionless parameter of HDE, which solely determines the evolution of HDE). The HDE model \cite{Li:2004rb} has been widely studied (see Ref.~\cite{hde} for some recent investigations; see Ref.~\cite{Wang:2016och} for a recent review). A recent study~\cite{Xu:2016grp} on comparing popular dark energy models using the latest observations shows that the HDE model is still a competitive candidate of dark energy among many models. 

I used the Planck 2015 temperature and polarization data, in combination with other low-redshift observations, including the baryon acoustic oscillations (BAO), type Ia supernovae (SN), and Hubble constant ($H_0$) measurement, as well as the Planck lensing measurements, to do the cosmological fits. I found that, compared to $\Lambda$CDM, once a dynamical dark energy is considered, the degeneracy between $\sum m_\nu$ and $H_0$ will be changed, i.e., in the $\Lambda$CDM model, $\sum m_\nu$ is anti-correlated with $H_0$, but in the $w$CDM model and the HDE model, $\sum m_\nu$ becomes positively correlated with $H_0$. I also showed that, compared to $\Lambda$CDM, in the $w$CDM model the limit on $\sum m_\nu$ becomes much looser, but in the HDE model the limit becomes much tighter. Using the data combination of Planck+BAO+SN+$H_0$+lensing, I obtained $\sum m_\nu<0.197$ eV for $\Lambda$CDM, $\sum m_\nu<0.304$ eV for $w$CDM, and $\sum m_\nu<0.113$ eV for HDE. Note that all the upper limit values for the total neutrino mass $\sum m_\nu$ quoted in this paper refer to the 95\% confidence level (CL). Therefore, we find that an extremely stringent upper limit, $\sum m_\nu<0.113$ eV, is obtained in the HDE model, which is much tighter than that obtained in the $\Lambda$CDM model.

This result has important implications. We know that, for IH of neutrino mass, the lower limit is approximately 0.1 eV, thus the upper limit obtained in this work is almost equal to the lower limit, implying that in the HDE model the IH seems to be nearly excluded. On the other hand, if the future neutrino oscillation experiments, such as the JUNO (Jiangmen Underground Neutrino Observatory) experiment, can successfully give the result of the neutrino mass ordering, and if the IH is the final answer, then according to the result of my investigation \cite{Zhang:2015uhk}, the HDE would be excluded by the neutrino mass ordering experiment. This expectation is very tantalizing because the neutrino oscillation experiments would potentially offer a possible falsifying scheme for the HDE model that currently is still a competitive candidate of dark energy.

In a further study \cite{Zhao:2016ecj}, collaborators and I also considered the Chevallier-Polarski-Linder (CPL) parametrization of dark energy, $w(a)=w_0+w_a(1-a)$, which has two more parameters than $\Lambda$CDM. We found that, compared to the $w$CDM model, the CPL model leads to a much larger upper limit of $\sum m_\nu$. We showed that, in these models, a phantom dark energy ($w<-1$) or an early phantom dark energy (i.e., the quintom evolving from $w<-1$ to $w>-1$) is slightly more favored by current observations, which leads to the fact that in both $w$CDM and CPL models a larger upper limit of $\sum m_\nu$ is obtained.  While in Ref.~\cite{Zhang:2015uhk} I showed that, in the HDE model, an early quintessence dark energy with $c<1$ (i.e., the quintom evolving from $w>-1$ to $w<-1$) is favored, and thus a smaller upper limit of $\sum m_\nu$, compared to $\Lambda$CDM, is obtained. In addition, in Ref.~\cite{Chen:2015oga}, the authors investigated the case of a tracking quintessence model with an inverse power-law potential, $V(\phi)\propto \phi^{-\alpha}$ ($\alpha>0$). They found that, in this model (i.e., the freezing quintessence evolving from $w>-1$ to $w\rightarrow -1$), a smaller upper limit of $\sum m_\nu$ is more favored, compared to $\Lambda$CDM (they obtained  $\sum m_\nu<0.262$ eV for the quintessence model and $\sum m_\nu<0.293$ eV for the $\Lambda$CDM model, under the same condition). Summarizing the results in both the HDE model and the tracking quintessence model, we conclude that, once $w$ evolves from a larger value to a smaller value, a smaller upper limit of $\sum m_\nu$, compared to $\Lambda$CDM, will be obtained.

Furthermore, collaborators and I \cite{Wang:2016tsz} began to take the mass splitting between the three active neutrinos into account in dynamical dark energy models. We also compared the models of $\Lambda$CDM, $w$CDM, and HDE, in this work. We showed that, the conclusion in Ref.~\cite{Zhang:2015uhk} is unchanged, i.e., the upper limit on $\sum m_\nu$ becomes much looser in the $w$CDM model but much tighter in the HDE model. Using the data combination of Planck+BAO+SN+$H_0$+lensing (where, compared to Ref.~\cite{Zhang:2015uhk}, BOSS DR11 is replaced with DR12 in BAO data, and the latest $H_0$ measurement is used), we obtained the upper limit $\sum m_\nu<0.105$ eV for the case of degenerate hierarchy (DH) of neutrinos in the HDE model, which is comparable to the lower limit of $\sum m_\nu$ for three inverted hierarchical neutrinos. This constraint is more stringent than that in Ref.~\cite{Zhang:2015uhk}. To our knowledge, this is perhaps the most stringent upper limit on the total mass of three degenerate neutrinos by far. We are on the verge of diagnosing the neutrino mass hierarchy through cosmological observations. We also found that, for all the dark energy models considered in this work, the minimal $\chi^2$ in the NH case is slightly smaller than that in the IH case. Thus, the NH case fits the current observations better than the IH one. But, actually, the difference $\Delta\chi^2$ is not yet significant enough to distinguish the neutrino mass hierarchy.  

More recently, the impacts of the possible coupling between dark energy and dark matter on constraining neutrino mass were also considered in Ref.~\cite{Guo:2017hea}. To avoid the large-scale instability problem in interacting dark energy models, the parametrized post-Friedmann (PPF) approach \cite{ppf} was employed in this study. To explicitly show the impacts from a coupling, the scenario of vacuum energy interacting with cold dark matter was investigated in detail. It was shown in Ref.~\cite{Guo:2017hea} that when the $Q=\beta H\rho_{\rm vac}$ model is considered, a smaller upper limit on $\sum m_\nu$ will be obtained, compared to $\Lambda$CDM under the same condition. 


In summary, through the brief review of the recent results of constraining neutrino mass in dynamical dark energy models using cosmological observations, we can get some conclusions. (i) In dynamical dark energy models, compared to $\Lambda$CDM, the upper limit of $\sum m_\nu$ can become larger and can also become smaller. We find that, in the cases of phantom and early phantom (i.e., the quintom evolving from $w<-1$ to $w>-1$), the constraint on $\sum m_\nu$ becomes looser; but in the cases of quintessence and early quintessence (i.e., the quintom evolving from $w>-1$ to $w<-1$), the constraint on $\sum m_\nu$ becomes tighter. (ii) In the HDE model, we can get the tightest constraint on $\sum m_\nu$, i.e., $\sum m_\nu<0.105$ eV (for the DH case of neutrinos), which is almost equal to the lower limit of $\sum m_\nu$ of IH case. Thus, it is of great interest to find that the future neutrino oscillation experiments would potentially offer a possible falsifying scheme for the HDE model. (iii) The mass splitting of neutrinos can influence the cosmological fits. We find that the NH case fits the current observations slightly better than the IH case, although the difference of $\chi^2$ of the two cases is not yet big enough to definitely distinguish the neutrino mass hierarchy. These statements need to be further carefully checked. To determine the neutrino mass and distinguish the mass hierarchy of three active neutrinos, more highly accurate observational data are needed.

\begin{acknowledgments}
This work was supported by the National Natural Science Foundation of China (Grants No.~11522540 and No.~11690021), the Top-Notch Young Talents Program of China, and the Provincial Department of Education of Liaoning (Grant No.~L2012087).

\end{acknowledgments}

\end{document}